\documentclass[reprint,showpacs,amsmath,amssymb,aps,prb]{revtex4-1}

\usepackage{graphicx}
\usepackage{graphics}
\usepackage{amsmath}
\usepackage{dcolumn}
\usepackage{bm}
\usepackage{subfig}

\begin{document}


\title{Floquet Weyl fermions in circularly-polarised-light-irradiated three-dimensional stacked graphene systems}

\author{Jin-Yu Zou}

\author{Bang-Gui Liu}%
 \email{bgliu@iphy.ac.cn}

\affiliation{Beijing National Laboratory for Condensed Matter Physics, Institute of Physics, Chinese Academy of Sciences, Beijing 100190, China.}

\date{\today}

\begin{abstract}
Using Floquet theory, we illustrate that Floquet Weyl fermions can be created in circularly-polarised-light-irradiated three-dimensional stacked graphene systems. One or two semi-Dirac points can be formed due to overlapping of Floquet sub-bands. Each pair of Weyl points have a two-component semi-Dirac point parent, instead of a four-component Dirac point parent. Decreasing the light frequency will make the Weyl points move in the momentum space, and the Weyl points can approach to the Dirac points when the frequency becomes very small. The frequency-amplitude phase diagram is worked out. It is shown that there exist Fermi arcs in the surface Brillouin zones in circularly-polarised-light-irradiated semi-infinitely-stacked and finitely-multilayered graphene systems. The Floquet Weyl points emerging due to the overlap of Floquet sub-bands provide a new platform to study Weyl fermions.
\end{abstract}



\maketitle


\section{Introduction}

As an important fundamental particle, Weyl fermion has been studied in hight-energy physics for a long time, but not been found in nature yet\cite{Weyl1929}. Nevertheless, in condensed matter physics, Weyl fermion can emerge as a quasiparticle in the so called Weyl semimetal (WSM)\cite{Nielsen1983,Murakami2007}. This phenomenon is analogue to Dirac points in graphene with band crossing and linear dispersion. Recently, the Weyl semimetal, with its novel properties, triggered enormous research activities\cite{Nielsen1983,Murakami2007,Burkov2011,Wan2011,Balents2011,Xu2011,Yang2011,Wang2012,Son2013,Wang2013,Vafek2014,Liu2014,Liu2014a,Borisenko2014,Neupane2014,Xu2015a,add1,add2}, especially the milestone experimental discovery of the Weyl fermions\cite{Xu2015,Lu2015,Lv2015}. Contrasting with the Dirac points in 2D graphene, Weyl fermions in the 3D semimetal are topologically stable, because the Hamiltonian uses up all three Pauli matrixes so that any perturbation can just move the Weyl points until meeting and annihilating with others with opposite topological numbers\cite{Murakami2007}. Topological property of WSM protects its surface gapless states which appear in the form of Fermi arc\cite{Wan2011,Balents2011,Xu2011}. The Fermi arc connects the projections in the surface Brillouin zone of opposite Weyl points. WSM also exhibits many other interesting phenomena, such as chiral anomaly\cite{Nielsen1983,Son2013}, anomalous Hall effect\cite{Xu2011}, negative magnetoresistance  under parallel electric and magnetic fields and others\cite{Fukushima2008,Son2013}.

In quantum field theory, a massless four-component Dirac fermion can be reduced to two two-component Weyl fermions, which enables one to define a new good quantum number respectively according to their chirality\cite{Pal2011}. In condensed matter physics, Weyl fermions can be created in a similar way\cite{Burkov2011,Wan2011,Balents2011,Yang2011,Xu2011,Wang2012,Son2013,Wang2013,Vafek2014,Wang2014}. One should find a matter with four-component massless Dirac point band structure, which means that the energy bands should touch with each other at the Dirac point with quadruple degeneracy and disperse linearly. The degeneracy of the Weyl points is protected by time-reversal symmetry and inversion symmetry. Breaking any of them will split them and create two Weyl points. In this sense, the Weyl point has a four-component Dirac parent.

Here, however, we will create Weyl fermions with a two-component semi-Dirac parent in a 3D lattice, a stacked graphene system irradiated by a circularly polarized light. Such a system can be solved using Floquet theory. When the frequency of light is much large compared to the hopping parameters of lattice, Floquet sub-bands are far from each other. With decreasing the frequency, sub-bands can get closer and closer to touch and overlap, and then separate and so forth\cite{Gomez-Leon2013,Kundu2014,Quelle2015,Lindner2011,Sentef2015}. Our investigation shows that  with the frequency decreasing, each time when the $\pm1$ Floquet sub-bands touch, a semi-Dirac point will be created and then split into two Weyl points with opposite chirality, producing two Weyl fermions. A frequency-amplitude phase diagram is completed. It is interesting that there can be one or two pairs of such Weyl fermions in the regime of low light amplitude. It is also shown that a Fermi arc can be created when a good surface is made. More detailed results will be presented in the following.

\section{Model and Floquet theory}

We define $\vec{\delta}_1$ to point to the $x$ direction and $\vec{a}_1$ the $y$ direction, with the $z$ direction being perpendicular to the paper plane. The three-dimensional lattice model consists of infinite graphene layers, each of which can be considered to be translated from its nearest graphene layer by the vector $\vec{a}_3$ in the $x-z$ plane\cite{Korhonen2015}, as  demonstrated in Fig. 1. We consider only the nearest-neighbor (NN) hopping constants within each layer and between the nearest layers. The light travels in $y$ direction, with its vector potential being in the $x-z$ plane:
\begin{equation}\label{4}
  \mathbf{A}(t)=A_0(\sin{wt},0,\cos{wt})
\end{equation}
By the Perierls substitution, $\mathbf{k}\rightarrow\tilde{\mathbf{k}}=\mathbf{k}+\mathbf{A}(t)$, we can obtain the time-dependent Hamiltonian,
\begin{equation}\label{5}
  H(\mathbf{k},t)=\left(
                    \begin{array}{cc}
                      0 & h_{12}(\mathbf{k},t) \\
                      h_{21}(\mathbf{k},t) & 0 \\
                    \end{array}
                  \right)
\end{equation}
where $h_{12}(\mathbf{k},t)$ is defined as $\gamma\sum_ne^{i(\mathbf{k}+\mathbf{A}(t))\cdot\mathbf{\delta}_n} +\eta e^{i(\mathbf{k}+\mathbf{A}(t))\cdot\mathbf{d}}$, and $h_{21}(\mathbf{k},t)$ is the complex conjugate of $h_{12}(\mathbf{k},t)$.
$\gamma$ is used to denote the NN hopping parameter within the graphene layer and $\eta$ the hopping parameter between the nearest graphene layers. The intra-layer NN distance is denoted as $a$ and the inter-layer distance in equivalent to $d\sin{\theta}$. We define both $\eta$ and $\gamma$ to be positive without losing any generality.

Without applying the light, we can obtain the 2D standard graphene if letting $\eta=0$, and a nonzero $\eta$ destroys the $C_3$ symmetry but reserves inversion symmetry. Therefore, the band structure is gapless when $\eta$ is less than $\gamma$. The gapless $\mathbf{k}$ points are nothing but special Dirac points defined with only two Pauli matrixes, pretty like Dirac points in the 2D graphene. They can survive even when the light turns on.

\begin{figure}[!tbp]
  \centering
  \includegraphics[clip, width=8cm]{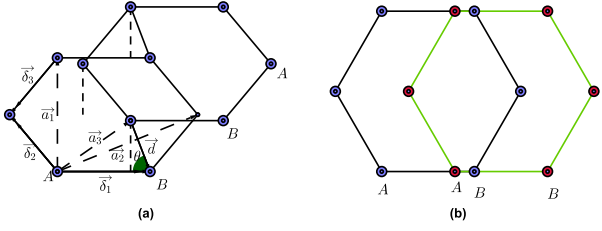}\\
  \caption{A schematic of the three-dimensional stacked graphene model which can be built by stacking infinite graphene layers. (a) shows the relation between the two nearest graphenes. The inter-layer distance is eqivalent to $d\sin\theta$. The sliding displacement between the nearest graphene layers, with the length of $a-d\cos\theta$, is along the $x$ axis, which becomes clear by using the top view (b). }\label{model}
\end{figure}

We use the Floquet theory to study the time-dependent non-equilibrium systems\cite{Sambe1973,Hemmerich2010} because the Hamiltonian (2) is periodic in time, $H(\mathbf{k},t)=H(\mathbf{k},t+T)$, where the time period is $T=2\pi/\omega$. In terms of Floquet-Bloch ansatz\cite{Platero2004}, the eigenstates can be written as $|\Psi_{\alpha,\mathbf{k}}(t)\rangle=e^{-i\varepsilon_{\alpha,\mathbf{k}}t}|u_{\alpha,\mathbf{k}}(t)\rangle$, where $\varepsilon_{\alpha,\mathbf{k}}$ stands for the quasi-energy of the Floquet state, $\alpha$ is the band index, and $|u_{\alpha,\mathbf{k}}(t)\rangle$ is periodic in $t$. Substituting $|\Psi_{\alpha,\mathbf{k}}(t)\rangle$ into the Sch\"{o}dinger equation leads to
\begin{equation}\label{1}
  (H(\mathbf{k},t)-i\partial_t)|u_{\alpha,\mathbf{k}}(t)\rangle=\varepsilon_{\alpha,\mathbf{k}}|u_{\alpha,\mathbf{k}}(t)\rangle
\end{equation}
We can define the Floquet Hamiltonian as
\begin{equation}\label{2}
  H_F=H(\mathbf{k},t)-i\partial_t
\end{equation}

Defining $|u_{\alpha,\mathbf{k},n}\rangle$ to be the Fourier series of the periodic Floquet state $|u_{\alpha,\mathbf{k}}(t)\rangle$, and applying $H_F$ on the basis $\{|u_{\alpha,\mathbf{k},n}\rangle\}$, we can express the operator $H_F$ explicitly in the composed Hilbert space $\mathcal{S}=\mathcal{H}\otimes\mathcal{T}$\cite{Sambe1973}, with inner product $\langle\langle\cdots\rangle\rangle=\int_0^T\langle\cdots\rangle dt/T$, where $\mathcal{T}$ is spanned by the $T$-periodic  function. With the annihilation and creating operator $c_{\alpha,\mathbf{k}}(t)$ and $c_{\alpha,\mathbf{k}}^\dag(t)$ Fourier expanded in $c_{\alpha,\mathbf{k},n}$ and $c_{\alpha,\mathbf{k},n}^\dag$, $H_F$ can be expressed as a block matrix form with its block elements\cite{Quelle2015}
\begin{equation}\label{3}
\begin{split}
  (H_F)_{n,m}&=nw\delta_{n,m}+H_{m-n}\\
  H_{m-n}&=\frac{1}{T}\int_0^TH(k,t)e^{iw(m-n)t}dt
\end{split}
\end{equation}
Obviously, $H_{m-n}$ are the Fourier modes of time periodic $H(k,t)$.
Without much effort, one can write the Floquet Hamiltonian as the block matrix form:
\begin{equation}\label{6}
  H_F(\mathbf{k})=\left(
                     \begin{array}{ccccc}
                       \ddots & \ddots & \ddots & \ddots & \ddots \\
                       \ddots & H_0+w & H_1 & H_2 & \ddots \\
                       \ddots & H_{-1} & H_0 & H_1 & \ddots \\
                       \ddots & H_{-2} & H_{-1} & H_0-w & \ddots \\
                       \ddots & \ddots & \ddots & \ddots & \ddots \\
                     \end{array}
                   \right)\\
\end{equation}
The $2\times 2$ matrix block $H_n$ can be expressed as
\begin{equation}\label{7}
  H_n=\frac{1}{T}\int_0^TdtH(\mathbf{k},t)e^{inwt}=\left(
         \begin{array}{cc}
           0 & (H_n)_{12} \\
          (H_n)_{21} & 0 \\
         \end{array}
       \right)
\end{equation}
Defining $v_i=e^{i\mathbf{k}\cdot\delta_i}$, we can write the matrix elements in Eq. (7) as
\begin{widetext}
\begin{equation}\label{8}
\left\{
\begin{split}
     (H_n)_{12}&=\gamma\{J_{-n}(A_0a)v_1+J_n(\frac{A_0a}{2})(v_2+v_3)\}+\eta e^{in\theta}J_{-n}(A_0d)e^{i\mathbf{k}\cdot\mathbf{d}}\\
     (H_n)_{21}&=\gamma\{J_{n}(A_0a)v_1^{*}+J_{-n}(\frac{A_0a}{2})(v_2^{*}+v_3^{*})\}+\eta e^{in\theta}J_{n}(A_0d)e^{-i\mathbf{k}\cdot\mathbf{d}}
\end{split}
\right.
\end{equation}
\end{widetext}
where $J_n$ is the Bessel function of order $n$.

\begin{figure*}[htbf]
  \centering
  \subfloat[$A_0a=1$,$w=3.5\gamma$]{
  \includegraphics[clip, width=.3 \textwidth, height=.28\textheight]{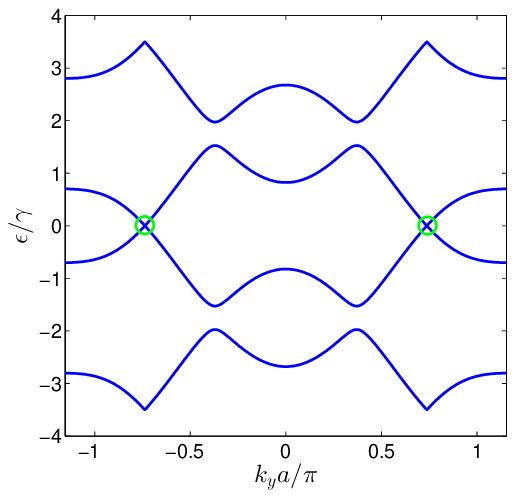}}
  \subfloat[$A_0a=1$,$w=2.3\gamma$]{
  \includegraphics[clip, width=.3 \textwidth, height=.28\textheight]{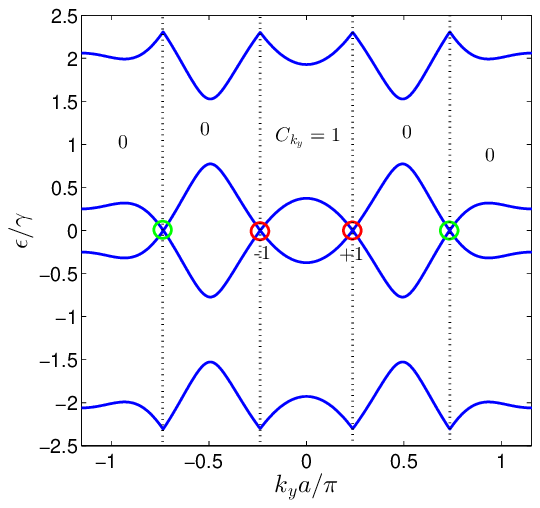}}
  \subfloat[$A_0a=1$,$w=1.5\gamma$]{
  \includegraphics[clip, width=.3 \textwidth, height=.28\textheight]{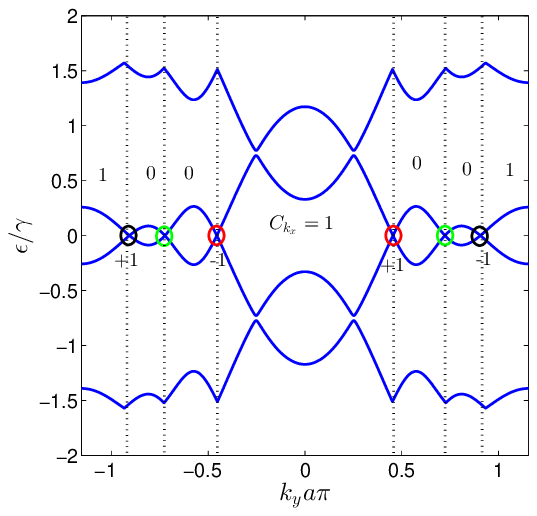}}\\
  \subfloat[$A_0a=2$,$w=3.5\gamma$]{
  \includegraphics[clip, width=.3 \textwidth, height=.28\textheight]{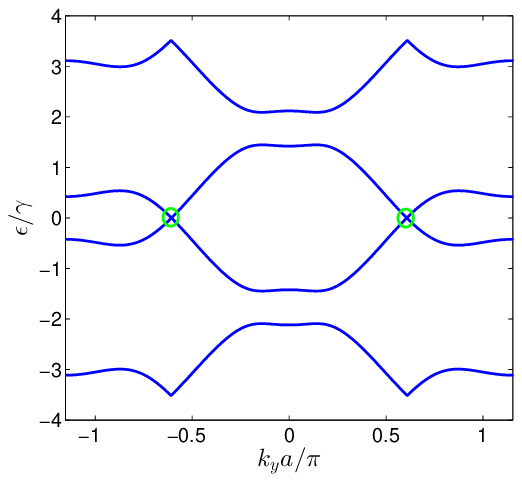}}
  \subfloat[$A_0a=2$,$w=2\gamma$]{
  \includegraphics[clip, width=.3 \textwidth, height=.28\textheight]{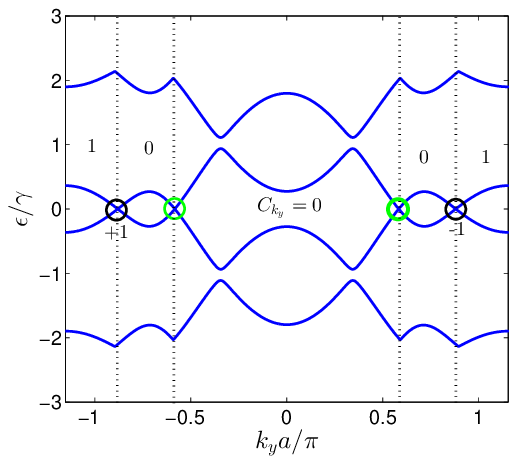}}
  \subfloat[$A_0a=2$,$w=1.3\gamma$]{
  \includegraphics[clip, width=.3 \textwidth, height=.28\textheight]{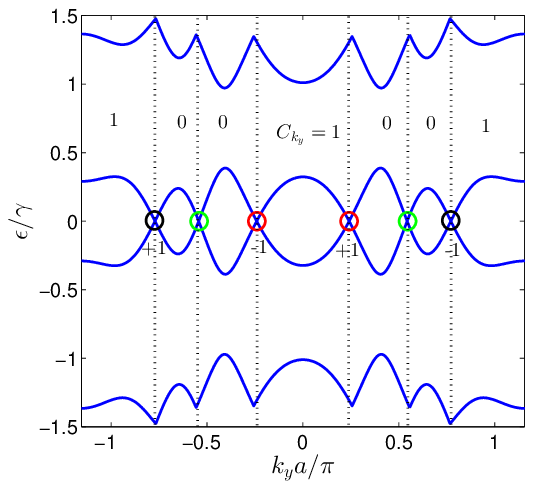}}\\
  \caption{The band structure along the $k_y$ axis, with $k_x=k_z=0$. For the $A_0a$ values, there always exist two gapless Dirac points (circled by green circles). For different frequency, there can be one/two pairs of Weyl points astride $\mathbf{k}_0$ (circled by red circles) or/and $\mathbf{k}_1$ (circled by black circles). $H_F$ matrix is truncated at $m=\pm2$, and other parameters are set as $d=2a$, $\eta=\gamma/7$, and $\theta=4\pi/9$.}\label{kx0kz0bands}
\end{figure*}

\section{Weyl Semimetal and Weyl fermions}

\subsection{Dirac and semi-Dirac points}

Generally speaking, there are infinite Floquet energy sub-bands labeled with $m=0,\pm1,\pm2,\cdots$.
To begin with, we consider the high-frequency limit, where the frequency is much larger than the hopping parameters ($w\gg\gamma,\eta$). In this regime, the overlapping between different Floquet sub-bands is negligible. As a result, the Floquet bands near the Fermi energy are determined by $H_0$, and the bands are gapless at points $\mathbf{k}^D_{\pm}=(0,\pm k_{y0},0)$ if the condition
\begin{equation}
\gamma(J_0(A_0a)+2J_0(\frac{1}{2}A_0a)\cos{(\frac{\sqrt{3}}{2}ak_{y0})})+\eta J_0(A_0d)=0
\end{equation}
is satisfied. These points, appearing in pair, are similar to usual Dirac points in the case of graphene, and are presented as the crossing points circled with green color in Fig. 2. Their low-energy effective Hamiltonian can be written as
\begin{equation}\label{ky0}
\begin{split}
  H^{\rm D}_{\rm eff}=&h_1\sigma_1+h_2\sigma_2\\
  h_1=&-\sqrt{3}a\gamma J_0(\frac{1}{2}A_0a)\sin{(\pm k_{y0}\frac{\sqrt{3}}{2}a)}k_y\\
  h_2=&a\gamma(J_0(\frac{1}{2}A_0a)\cos{(k_{y0}\frac{\sqrt{3}}{2}a)}-J_0(A_0a))k_x\\
       &-\eta J_0(dA_0)\mathbf{k}\cdot\mathbf{d}
\end{split}
\end{equation}
We use usual definition for the Pauli matrices: $\sigma_1=\sigma_x$, $\sigma_2=\sigma_y$, and $\sigma_3=\sigma_z$. These mean that in this frequency regime the system is a Dirac semimetal. When the light strength increases, the two gapless points will move towards the origin of the momentum space and will finally disappear at the origin, opening a semiconductor gap and causing the Dirac semimetal to become an ordinary insulator.

When the frequency decreases further, there will be some overlapping between different Floquet sub-bands.
With the frequency decreasing, at first, the lower one of the $m=1$ sub-bands will intersect with the upper one of the $m=0$ sub-bands at $\varepsilon=w/2$ and meanwhile the upper one of the $m=-1$ bands will intersect with the lower one of the $m=0$ bands at $\varepsilon=-w/2$. As a rule, the near sub-bands whose $m$ values differ by $\Delta m=\pm1$ will intersect. It should be pointed out that the perturbation introduced by $H_{\pm1}$, that stands for the coupling of one-photon stressed sub-bands and zero sub-bands by absorbing and emitting one photon, can open energy gaps at the band crossing points. Here, we will concentrate on the band touching between the $m=\pm1$ sub-bands at $\varepsilon=0$. There is one touching point at $\mathbf{k}_0=(0,0,0)$) when the condition
\begin{equation}\label{111}
w=\gamma(J_0(A_0a)+2J_0(\frac{1}{2}A_0a))+\eta J_0(A_0d)
\end{equation}
is satisfied. Another touching point can appear at $\mathbf{k}_1=(0,\frac{2}{\sqrt{3}a}\pi,0)$) when the condition
\begin{equation}\label{222}
w=\gamma(J_0(A_0a)-2J_0(\frac{1}{2}A_0a))+\eta J_0(A_0d)
\end{equation}
is satisfied.
Unfortunately, however, the $w$ value can be changed a little by the perturbation of $H_{\pm1}$ when the $m=\pm1$ sub-bands touch with each other. Being different from the $\Delta m=\pm1$ sub-bands touching, the perturbation of the $H_{\pm2}$ block will be diagonal between the $\pm1$ sub-bands when we diagonalize the $H_0$ block along the $k_y$ axis. That means the band touching points can be moved but cannot be removed by the perturbation. This trend can be seen in Fig. 2.

The $k\cdot p$ perturbation theory can be used to study the band structure near the two touching points at $\mathbf{k}_0=(0,0,0)$ and $\mathbf{k}_1=(0,\frac{2}{\sqrt{3}a}\pi,0)$, respectively. It can be easily found out that they are semi-Dirac points with effective hamiltonian
\begin{equation}\label{9}
\begin{split}
    H^{\rm SD}_{\rm eff}=&\sum^3_{i=1} h_i(\mathbf{k})\sigma_i\\
    h_1(\mathbf{k})=&-2\eta\sin{2\theta}J_2(A_0d)\mathbf{k}\cdot\mathbf{d}\\
    h_2(\mathbf{k})=&-2\gamma[J_2(A_0a)\mp J_2(\frac{1}{2}A_0a)]k_xa\\
                    &-2\eta\cos{2\theta}J_2(A_0d)\mathbf{k}\cdot\mathbf{d}\\
    h_3(\mathbf{k})=&-\gamma \{J_0(A_0a)(\mathbf{k}\cdot\mathbf{a}_1)^2\pm J_0(\frac{1}{2}A_0a)[(\mathbf{k}\cdot\mathbf{a}_2)^2\\
                    &+(\mathbf{k}\cdot\mathbf{a}_3)^2]\}-\eta J_0(A_0d)(\mathbf{k}\cdot\mathbf{d})^2
\end{split}
\end{equation}
where $\mathbf{k}$ is defined in the vicinity of $\mathbf{k}_0$ or $\mathbf{k}_1$. The upper signal in $\mp$ and $\pm$ in Eq. (13) is for $\mathbf{k}_0$, the lower one for $\mathbf{k}_1$. These semi-Dirac points are similar to those in the 2D graphene formed by making the two Dirac points along the $y$ axis to meet when the $C_3$ symmetry was broken\cite{Bernevig2013}, but in that case, the graphene's band structure is opened a gap after the two semi-Dirac points is merged.

\subsection{Nontrivial Weyl points and Weyl fermions}

The energy gap between the $m=\pm1$ sub-bands is closed upon the semi-Dirac point is created. With the frequency decreases further, the semi-Dirac point will split into two Weyl points. The two Weyl points are located at $\mathbf{k}^c_{\pm}=(0,\pm k_{yc},0)$, where $k_{yc}$ is determined by the equation
\begin{equation}
w=\gamma(J_0(A_0a)+2J_0(\frac{1}{2}A_0a)\cos{(\frac{\sqrt{3}}{2}ak_{yc})})+\eta J_0(A_0d).
\end{equation}
Similarly the effective Hamiltonian can be obtained by the $k\cdot p$ perturbation theory, reading
\begin{equation}\label{10}
\begin{split}
       H^{\rm W}_{\rm eff}=& \sum^3_{i,j=1}v_{ij}k_i\sigma_j\\
       v_{11}=&-2\eta\sin{2\theta}J_2(A_0d)d\cos{\theta}\\
       v_{12}=&2a\gamma[J_2(\frac{1}{2}A_0a)\cos{(k_{yc}\frac{\sqrt{3}}{2}a)}\\
              &-J_2(A_0a)]-2\eta\cos{2\theta}J_2(A_0d)d\cos{\theta}\\
       v_{23}=&2\sqrt{3}a\gamma J_2(\frac{1}{2}A_0a)\sin{(\pm k_{yc}\frac{\sqrt{3}}{2}a)}\\
       v_{31}=&2\eta\sin{2\theta}J_2(A_0d)d\sin{\theta}\\
       v_{32}=&2\eta\cos{2\theta}J_2(A_0d)d\sin{\theta}\\
       v_{13}=&v_{21}=v_{22}=v_{33}=0
\end{split}
\end{equation}
It is interesting that the low-energy Hamiltonians for the Weyl point pairs astride both $\mathbf{k}_0$ and $\mathbf{k}_1$ have the same form. If we let $k_{yc}$ be equal to either $0$ or $\frac{2}{\sqrt{3}a}\pi$, the $v_{23}$ term will disappear and the quadric term will return, resuming a semi-Dirac point again.

\begin{figure}[!htbp]
  \centering
  \includegraphics[clip, width=8cm]{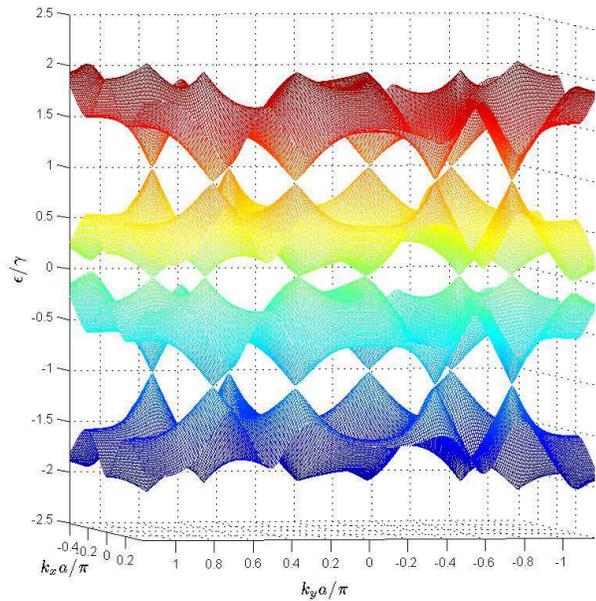}\\
  \caption{The band structures of the $m=0$ and $m=\pm1$ sub-bands in the $k_z=0$ plane, with parameters $\eta=\gamma/7$, $A_0a=1.5$, $w=2\gamma$, $d=2a$, and $\theta=4\pi/9$. Two pairs of Weyl points are symmetrical astride $\mathbf{k}_0$ and $\mathbf{k}_1$, respectively.}\label{band structure}
\end{figure}

We can calculate the Weyl number using the definition
\begin{equation}\label{C}
\begin{split}
  C=&{\rm sgn}[\det[v]]\\
   =&{\rm sgn}[4\sqrt{3}a^2\gamma J_2(\frac{1}{2}A_0a)\sin{(k_{yc}\frac{\sqrt{3}}{2}a)}*\\
    &(J_2(\frac{1}{2}A_0a)\cos{(k_{yc}\frac{\sqrt{3}}{2}a)}-J_2(A_0a))]
\end{split}
\end{equation}
It can be proved that the two Weyl points possess opposite Weyl number or chirality. The Weyl points created by the band crossing of the $m=\pm1$ sub-bands, with their Weyl numbers labeled, are shown in Fig. 3.

\subsection{Phase diagram}

When the frequency decreases, the distance between the two Weyl points in a pair become more separated and $k_{yc}$ can approach $k_{y0}$ but cannot exceed it. Thus, in this frequency regime, the system will stay in a WSM phase, with one pair of Weyl points or two if one or both of the semi-Dirac points (responding to $\mathbf{k}_0$ and $\mathbf{k}_1$) were created. On the other hand, for the very strong light amplitude, we should have an ordinary insulator phase in the regime of high frequency. In this strong light regime, there can be a pair of Weyl points astride $\mathbf{k}_1$ when the frequency becomes enough low. The complete phase diagram is presented in Fig. 4. We have six phases: semi-metal (SM), Weyl semi-metal with one pair of Weyl fermions astride $\mathbf{k}_0$ (WSM1-A), Weyl semi-metal with one pair of Weyl fermions astride $\mathbf{k}_1$ (WSM1-B), Weyl semi-metal with two pair of the Weyl fermions (WSM2), ordinary insulator (I), and Weyl semi-metal (WSM). There are Dirac points in the former four phases, but there exists no Dirac points in the I and WSM phases.

\begin{figure}[!htbp]
  \centering
  \includegraphics[clip, width=7cm]{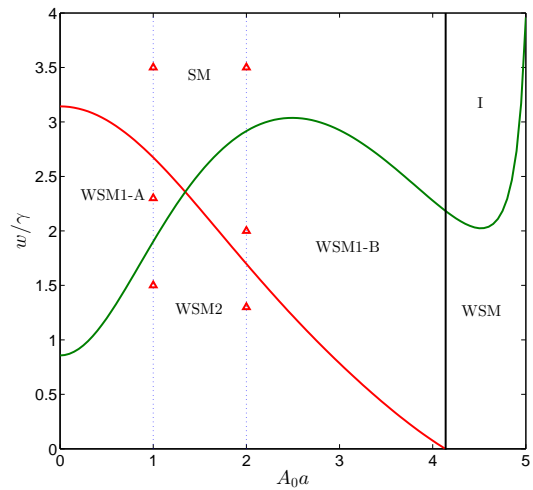}\\
  \caption{The phase diagram with experimentally achievable\cite{Korhonen2015} parameters: $d=2a$, $\eta=\gamma/7$, and $\eta=4\pi/9$. There are six phases: semi-metal (SM), Weyl semi-metal with one pair of Weyl fermions astride $\mathbf{k}_0$ (WSM1-A), Weyl semi-metal with one pair of Weyl fermions astride $\mathbf{k}_1$ (WSM1-B), Weyl semi-metal with two pair of the Weyl fermions (WSM2), ordinary insulator (I), and Weyl semi-metal (WSM). The six triangles show the ($\omega/\gamma$,$A_0a$) values for the band structures shown in Fig. 2.}\label{weyl phase}
\end{figure}

\subsection{Surface states and finite multilayers}

One of the most striking phenomena in a WSM phase is its Fermi arc. We can address it in a simple way. In our case, $H^{\rm W}_{\rm eff}(k_x,k_y,k_z)$ with a given $k_y$ can be considered as a 2D ($k_x$,$k_z$) Hamiltonian and one can calculate its $k_y$-dependent Chern number $C_{k_y}$. It is well known that for a 2D Hamiltonian one can change its Chern number by changing its parameters. In our case, the Chern number will change by
\begin{equation}
\Delta C=C_{k_y}-C_{k^\prime_y}
\end{equation}
when $k_y$  crosses a Weyl point where the mass term changes sign. It is easy to get
$C_{k_y}=1$ when $k_y$ is between the two Weyl points in each pair, and $C_{k_y}=0$ otherwise. Semi-Dirac points do not change the Chern number because the $H^{\rm SD}_{\rm eff}$ near it has no mass term. We already illustrate the Chern numbers in Fig. 2. Each nontrivial 2D Hamiltonian have edge states located at its boundary and these edge states make the Fermi arc which connects the projections of each pair of Weyl points in the surface BZ. Therefore, as long as the surface is parallel to the $k_y$ axis, a Fermi arc will emerge in the surface Brillouin zone.

It is clear that to experimentally realize the Floquet Weyl fermions, some appropriate multilayer structures are better than the infinitely-stacked graphene model. On the basis of the above surface construction, we can construct finite stacked graphene models under circularly-polarized light. Such circularly-polarised-light-irradiated finitely-stacked models can be realized experimentally. They can host Floquet Weyl fermions and Fermi arcs.

\section{Further discussion and concludion}

For experimentally realizing Floquet Weyl fermions in such graphene-based models, one has to prepare the circularly-polarized light of suitable frequency and amplitude.
Because the NN hopping constants ($\gamma\approx$2.8eV, $a\approx1.4$\AA) are very large, we usually need a high frequency (order of magnitude of $\sim 10^{16}$Hz) and a strong amplitude (order of magnitude of $\sim 10^8$V/m). As shown in the phase diagram, however, the Floquet Weyl fermions can also be realized when the frequency is low or the amplitude is weak. It should be pointed out that in the regime of low frequency, other Floquet sub-bands, such as those with $\Delta m=4,6,8$, will play roles in creating Floquet Weyl points at $\epsilon=0$, and therefore some complex situations may appear. Fortunately, in the regime of weak field amplitude, one can realize one or two pairs of Floquet Weyl fermions at experimentally achievable frequency levels.

In addition, some self-assembled graphene-like lattices of CdSe nanostructures \cite{Kalesaki2014,Boneschanscher2014,Quelle2015} can be used to realize such Floquet Weyl fermions. Because their hopping parameters are approximately two order of magnitude smaller compared to graphene and their lattice constant is one order of magnitude larger, much lower frequency and weaker amplitude are needed to realize Floquet Weyl fermions in good multilayer structures of such CdSe nanostructures\cite{Korhonen2015}.

In conclusion, we have proposed an interesting method to create Floquet Weyl fermions with a two-component semi-Dirac parent in a circularly-polarised-light-irradiated 3D stacked graphene system instead of a four-component Dirac parent. One or two semi-Dirac points can appear at $\mathbf{k_0}=(0,0,0)$ or/and $\mathbf{k_1}=(0,\frac{2}{\sqrt{3}a}\pi,0)$ in momentum space when the frequency $w$ of light is on the order of magnitude of hopping parameters $\gamma$ and $\eta$. Upon decreasing the light frequency, each semi-Dirac point will split into two symmetrical Weyl points with opposite chirality. Further decreasing the frequency will make the Weyl points move in the momentum space, and the Weyl points can approach to the Dirac points when the frequency becomes very small. The frequency-amplitude phase diagram has been worked out, indicating that the Floquet Weyl femions always appear in pair. Furthermore, it has been shown that there exist Fermi arcs in the surface Brillouin zones in circularly-polarised-light-irradiated semi-infinitely-stacked and finitely-multilayered graphene systems. These theoretical results can lead to a new platform to create Weyl fermions in graphene-based and similar systems.

\begin{acknowledgments}
This work is supported by Nature Science Foundation of China (Grant Nos. 11174359 and 11574366), by Chinese Department of Science and Technology (Grant No. 2012CB932302), and by the Strategic Priority Research
Program of the Chinese Academy of Sciences (Grant No. XDB07000000).
\end{acknowledgments}




\end{document}